# Three Remarks on the MSSM

S. Sultansoy

Deutsches Elektronen-Synchrotron DESY, Hamburg, Germany
Department of Physics, Faculty of Sciences, Ankara University, Turkey
Institute of Physics, Academy of Sciences, Baku, Azerbaijan

**Abstract**

First of all, "right" sneutrino can be the lightest supersymmetric particle. Then, the flavor democracy favors large value of the tanβ. Finally, the huge number of free parameters in the three families MSSM may be considered as an indication of SUSY at preonic level.

---

Electronic address: sultanov@mail.desy.de
            sultan@science.ankara.edu.tr

## 1. Introduction

The huge number of free parameters [1,2] in the three families MSSM leads to consideration of some simplified versions, such as the constrained MSSM (see [3] and references therein). In general, these simplifications ignore interfamily mixings and possible existence of right-handed neutrinos, and consequently their super-partners. As the result one avoid possible conflicts with experimental data on flavor violating processes, but at the same time we also lose very interesting possible phenomenology. Below we deal with the three families MSSM and do not consider possible R-parity violation, as well as GUT and SUGRA extensions.

## 2. Parameter Inflation

The are a number of arguments favoring the existence of right-handed neutrinos. First of all, in the framework of the SM $\nu_R$'s are counterparts of the right-handed components of the down-type quarks according to the quark-lepton symmetry. Then, almost all extensions of the SM, with the SU(5) GUT as a possible exception, naturally contain right-handed neutrinos. Moreover, observation of the neutrino oscillations provides the experimental confirmation for $\nu_R$. For these reasons, we consider three families MSSM including right-handed neutrinos. Therefore, we deal with the six species that constitute a family, rather than five considered in [2] (below we follow the notations used in this paper): $q$, $\bar{d}$, $\bar{u}$, $l$, $\bar{e}$ and $\bar{\nu}$, where $q$ and $l$ denote weak iso-doublets and the rest are iso-singlets.

The masses of the SM fermions and their super-partners are generated due to

$$L = L_{scalar} + L_{Yukawa} + L_{triscalar} \,. \tag{1}$$

The first term has the form

$$L_{scalar} = \sum_{A,i,j} m^2_{Aij} \tilde{A}^*_i \tilde{A}_j \,, \tag{2}$$

where A labels above-mentioned six species, the tilde labels sparticle and i,j=1,2,3 are family labels. Therefore, this part contain six 3×3 Hermitian mass matrices with six real parameters and three phases each. The Yukawa part is derived from the superpotential

$$W_{Yukawa} = \sum_{i,j} (q_i \lambda_{uij} \bar{u}_j H_u + q_i \lambda_{dij} \bar{d}_j H_d + l_i \lambda_{\nu ij} \bar{\nu}_j H_u + l_i \lambda_{eij} \bar{e}_j H_d) \,, \tag{3}$$

where four Yukawa matrices $\lambda$ are general 3×3 matrices with nine real parameters and nine phases each. Finally, the triscalar part (in order to avoid confusion with the second equation we introduce here the notations, which are slightly different from that in [2]) is given by

$$L_{triscalar} = \sum_{i,j} (\tilde{q}_i a_{uij} \tilde{\bar{u}}_j H_u + \tilde{q}_i a_{dij} \tilde{\bar{d}}_j H_d + \tilde{l}_i a_{\nu ij} \tilde{\bar{\nu}}_j H_u + \tilde{l}_i a_{eij} \tilde{\bar{e}}_j H_d) \times M \,, \tag{4}$$

where *a* are general 3×3 matrices and M is the some mass parameter. As the result we have 108 real parameters (masses and mixing angles) and 90 phases. However, part of them are unobservable because of $[U(3)]^6$ symmetry of the gauge sector, so 18 angles and 34 phases can be rotated out and remaining two phases correspond to baryon number in quark sector and general lepton number. Finally, matter sector of the MSSM (fundamental fermions and their superpartners) contains 90 observable real parameters, namely 36 masses and 54 mixing angles, and 56 phases.

It is clear that tremendous number of physical parameters, which makes meaningless any general analysis, at the same time gives the opportunity to get the predictions up to the experimental upper limits almost for all rare processes. For this reason, from the viewpoint of the MSSM the investigation of charmed particles (oscillations, CP-violation etc) becomes as interest as investigation of beauty particles, the search for $\tau \to \mu\gamma$ and $\tau \to e\gamma$ is as important as the search for $\mu \to e\gamma$ etc.

## 3. Flavor Democracy

Let me remind you the main assumptions of the flavor democracy or, in other words, democratic mass matrix hypothesis (in the frameworks of the n family SM):

i) Before the spontaneous symmetry breaking fermions with the same quantum numbers are indistinguishable. Therefore [4], Yukawa couplings are equal within each type of fermions: $\lambda_{uij}=\lambda_u$, $\lambda_{dij}=\lambda_d$, $\lambda_{lij}=\lambda_l$ and $\lambda_{\nu ij}=\lambda_\nu$.

ii) There is only one Higgs doublet, which gives Dirac masses to all four types of fermions. Therefore [5], Yukawa constants for different types of fermions should be nearly equal: $\lambda_u \approx \lambda_d \approx \lambda_l \approx \lambda_\nu \approx \lambda$.

The first statement result in n-1 massless particles and one massive particle with m=n$\lambda_F$ (F=u, d, l, $\nu$) for each type of the SM fermions. The masses of the first n-1 families, as well as observable interfamily mixings, are generated due to a small deviation from the full flavor democracy [6]. Taking into account the mass values for the third generation, the second statement leads to the assumption that the fourth SM family should exist. Alternatively, masses of up and down type fermions should be generated by the interaction with different Higgs doublets, as it takes place in the MSSM.

Finally, flavor democracy provides the opportunity to get the massless states as the superposition of initially massive particles. This property may be useful for the preonic models because, in general, the masses of composite objects are expected to be of the order of compositness scale.

## 4. Sneutrino as the LSP

It is straightforward to apply flavor democracy to the MSSM. For example, according the flavor democracy sneutrino mass matrix has the form

$$\begin{pmatrix} m_{LL}^2 & m_{LL}^2 & m_{LL}^2 & m_{LR}^2 & m_{LR}^2 & m_{LR}^2 \\ m_{LL}^2 & m_{LL}^2 & m_{LL}^2 & m_{LR}^2 & m_{LR}^2 & m_{LR}^2 \\ m_{LL}^2 & m_{LL}^2 & m_{LL}^2 & m_{LR}^2 & m_{LR}^2 & m_{LR}^2 \\ m_{RL}^2 & m_{RL}^2 & m_{RL}^2 & m_{RR}^2 & m_{RR}^2 & m_{RR}^2 \\ m_{RL}^2 & m_{RL}^2 & m_{RL}^2 & m_{RR}^2 & m_{RR}^2 & m_{RR}^2 \\ m_{RL}^2 & m_{RL}^2 & m_{RL}^2 & m_{RR}^2 & m_{RR}^2 & m_{RR}^2 \end{pmatrix}. \quad (5)$$

As the result we deal with four massless sneutrinos and two sneutrinos have the masses

$$m_{3,6}^2 = \frac{3}{2}\left( m_{LL}^2 + m_{RR}^2 \mp \sqrt{(m_{LL}^2 - m_{RR}^2)^2 + 4 m_{LR}^2 m_{RL}^2} \right). \quad (6)$$

The (small) masses of the rest four species can be generated due to violation of flavor democracy and/or radiatively. Including F- and D-term contributions, the elements of the matrix (5) have the following form [3]:

$$m_{LL}^2 = m_l^2 + (\lambda_\nu v_u)^2 + \frac{1}{2} m_Z^2 \cos 2\beta,$$
$$m_{LR}^2 = m_{RL}^2 = a_\nu (M v_u - \mu v_d),$$
$$m_{RR}^2 = m_{\tilde\nu}^2 + (\lambda_\nu v_u)^2,$$

where $v_u$ and $v_d$ are vacuum expectation values of the Higgs fields $H_u$ and $H_d$, $tan\beta = v_u/v_d$ and $\mu$ is the supersymmetry-conserving Higgs mass parameter.

In the recent paper [7] it is shown that LEP1 data leads to lower bound 44.6 GeV on the sneutrino masses and, consequently, in the framework of the constrained MSSM sneutrino cannot be the LSP. Let me present some notes on this statement. First one is rather technical. Presented lower bound is valid for the case of degenerated superpartners of left-handed neutrinos. For non-degenerate case it should be change to 43.6 GeV for lightest one. Then, according flavor democracy the constrained MSSM seems not to be so natural at all. The most important note is following: LEP1 data does not essentially constrain the masses of superpartners of the right-handed neutrinos if the LR mixings are sufficiently small. Indeed, the contribution of the "right" sneutrino to the invisible Z width is given by

$$\Delta\Gamma_{inv} = 0.5 \times |\delta|^2 \times \left[ 1 - \left( \frac{2 m_{\tilde\nu}}{m_Z} \right)^2 \right]^{3/2} \times \Gamma_\nu, \quad (7)$$

where $\delta$ denotes the "left" sneutrino fraction due to corresponding mixings and $\Gamma_\nu$=167 MeV. Therefore, the experimental value $\Delta\Gamma_{inv} \leq 2.0$ MeV leads to $|\delta| \leq 0.155$ for sufficiently light "right" sneutrino. If there are two light species one obtain $|\delta_1|^2 + |\delta_2|^2 \leq 0.024$. Therefore, "right" sneutrino still can be considered as the LSP both in constrained and unconstrained MSSM. This scenario may lead to very interesting phenomenology, however, this is beyond the scope of the present paper.

## 5. Flavor Democracy and tanβ

According to the first assumption of the flavor democracy, in the framework of the three families MSSM the masses of t- and b-quarks are as follows:

$$m_t = 3 \times \lambda_t \times v_u, \quad m_b = 3 \times \lambda_b \times v_d. \tag{8}$$

Application of the second assumption, namely $\lambda_t \approx \lambda_b$, immediately leads to relation

$$\tan \beta = \frac{v_u}{v_d} \approx \frac{m_t}{m_b}. \tag{9}$$

With $m_t \approx 174$ GeV and $m_b \approx 4.3$ GeV [8] we obtain tanβ≈40. More conservatively (taking into account the "running" of the quark masses etc.), flavor democracy favors the region of 30<tanβ<50 and lower values can be interpreted as an indication of the fourth MSSM family. In the last case one has

$$\tan \beta \approx \frac{m_{u_4}}{m_{d_4}} \approx 1 \tag{10}$$

in order to satisfy $\rho_{exp}$=0.9998±0.0008 [8].

## 6. SUSY and Preons

The huge number of arbitrary parameters naturally leads to assumption that SUSY should be realized at more fundamental, preonic or even pre-preonic level. Indeed, three families MSSM with Dirac neutrinos contains 160 observable parameters (146 from quark-squark and lepton-slepton sectors and the rest from gauge couplings etc [3]) which should be compared with 26 observable parameters in SM case. Today, there is not any realistic way to decrease essentially this number. The statements about a few parameters in the SUGRA express rather our wishes, because they practically ignore interfamily mixings. In principle, CKM-like mixings (rather than family replication) can be considered as the serious indication of the compositness, because the superposition of fundamental objects with different masses is forbidden in the framework of the quantum field theory. Besides, the proposition [9] about similarity of CKM mixings and squared mass differences in quark and squark sectors for explanation of experimental data on flavor violating processes seems quite natural if MSSM matter fields are composite. On the other hand, there is not any realistic preonic model, also. However, one can made some general predictions for preonic SUSY without consideration of specific models.

Composite models of leptons and quarks can be divided into two classes: fermion-scalar models and three-fermion models. In the first class SM fermions are composites of scalar preons, denoted by S, and fermion preons, denoted by F. In minimal variant $q,l=\{FS\}$ and each SM fermion (FS) with m≈0 has three partners:

$$scalar\ (\tilde{F}S)\ with\ m \approx M_{SUSY},$$
$$scalar\ (F\tilde{S})\ with\ m \approx M_{SUSY},$$
$$fermion\ (\tilde{F}\tilde{S})\ with\ m \approx 2M_{SUSY},$$

where $M_{SUSY}$ denotes the SUSY scale. In the second class SM fermions are composites of three fermions, $q,l = (F_1 F_2 F_3)$, and each of them has seven partners:

$$three\ scalars\ (\tilde{F}_1 F_2 F_3),\ (F_1 \tilde{F}_2 F_3)\ and\ (F_1 F_2 \tilde{F}_3)\ with\ m \approx M_{SUSY},$$
$$three\ fermions\ (\tilde{F}_1 \tilde{F}_2 F_3),\ (\tilde{F}_1 F_2 \tilde{F}_3)\ and\ (F_1 \tilde{F}_2 \tilde{F}_3)\ with\ m \approx 2M_{SUSY},$$
$$scalar\ (\tilde{F}_1 \tilde{F}_2 \tilde{F}_3)\ with\ m \approx 3M_{SUSY}.$$

Of course, mixings between quarks (leptons, squarks, sleptons) can drastically change the simple mass relations given above. Therefore, it is quite possible that the search for SUSY at future colliders will give rather surprising results.

Finally, let me note that even one family MSSM contains two observable mixing angles and two phases both in quark-squark [1] and lepton-slepton sectors. This fact may be useful for construction of the supersymmetric preonic models.

## 7. Conclusion

Today, the mass and mixing patterns of the MSSM, as well as the SM itself, are the most mysterious aspects of the particle physics. It seems that the general tendency is toward the idea of A. Salam and collaborators on pre-preonic SUSY model [10]. Let me finish by two notes on the rare processes mentioned in the Section 2:

i) a huge number of charmed particles which will be produced at HERA-B [11] provide the strong tool for detailed investigation of charm physics, and

ii) an operation of the TESLA at Z-resonance (TESLA-Z [12]) will give opportunity to improve by an orders the upper limits on the Br($\tau\to\mu\gamma$) and Br($\tau\to e\gamma$).

**Acknowledgements**
I am grateful to I. Ginzburg, M. Spira and P. Zerwas for useful discussions and valuable remarks. This work was made during my sabbatical visit to DESY. I would like to express my gratitude to DESY Directorate for invitation and hospitality.

**P.S.** According to the recent classification of H.E. Haber [13], the three family MSSM should be denoted as "MSSM>160" and the four family MSSM as the "MSSM>272". The meaning of the symbol ">" will be explained in forthcoming paper [14].

**P.P.S.** Professor P. Zerwas drew my attention to the paper [15], which contains the statement similar to (9). Unfortunately, their approach based on minimal S0(10) model is excluded by the recent experimental values of $\alpha_s$ and $m_t$ [8].